\documentclass[aps,prl,twocolumn,showpacs,nofootinbib,amsmath,amssymb,amsfonts]{revtex4-1}

\usepackage{graphicx}
\usepackage{epsfig,latexsym}

\usepackage[usenames]{color}
\RequirePackage{xspace} \allowdisplaybreaks

\usepackage{natbib}
\usepackage{url}
\usepackage{lineno}

\usepackage{bm}
\usepackage{dcolumn}
\usepackage{graphicx}
\usepackage{graphics}
\usepackage[latin1]{inputenc}
\usepackage{latexsym}
\usepackage{rotating}
\usepackage{xspace} 

\usepackage{ulem}
\usepackage{outlines}
\usepackage{enumitem}
\setenumerate[1]{label=\Roman*.}
\setenumerate[2]{label=\Alph*.}
\setenumerate[3]{label=\roman*.}
\setenumerate[4]{label=\alph*.}

\definecolor {darkgreen}{rgb}{0.2,0.7,0.2}

\begin{document}

\def\bef{\begin{figure}}
\def\eef{\end{figure}}

\newcommand{\nl}{\nonumber\\}

\newcommand{\ans}{ansatz }
\newcommand{\be}[1]{\begin{equation}\label{#1}}
\newcommand{\beq}{\begin{equation}}
\newcommand{\ee}{\end{equation}}
\newcommand{\beqn}[1]{\begin{eqnarray}\label{#1}}
\newcommand{\eeqn}{\end{eqnarray}}
\newcommand{\bd}{\begin{displaymath}}
\newcommand{\ed}{\end{displaymath}}
\newcommand{\mat}[4]{\left(\begin{array}{cc}{#1}&{#2}\\{#3}&{#4}
\end{array}\right)}
\newcommand{\matr}[9]{\left(\begin{array}{ccc}{#1}&{#2}&{#3}\\
{#4}&{#5}&{#6}\\{#7}&{#8}&{#9}\end{array}\right)}
\newcommand{\matrr}[6]{\left(\begin{array}{cc}{#1}&{#2}\\
{#3}&{#4}\\{#5}&{#6}\end{array}\right)}
\newcommand{\cvb}[3]{#1^{#2}_{#3}}
\def\lsim{\raise0.3ex\hbox{$\;<$\kern-0.75em\raise-1.1ex
e\hbox{$\sim\;$}}}
\def\gsim{\raise0.3ex\hbox{$\;>$\kern-0.75em\raise-1.1ex
\hbox{$\sim\;$}}}
\def\abs#1{\left| #1\right|}
\def\simlt{\mathrel{\lower2.5pt\vbox{\lineskip=0pt\baselineskip=0pt
           \hbox{$<$}\hbox{$\sim$}}}}
\def\simgt{\mathrel{\lower2.5pt\vbox{\lineskip=0pt\baselineskip=0pt
           \hbox{$>$}\hbox{$\sim$}}}}
\def\unity{{\hbox{1\kern-.8mm l}}}
\newcommand{\eps}{\varepsilon}
\def\ep{\epsilon}
\def\ga{\gamma}
\def\Ga{\Gamma}
\def\om{\omega}
\def\omp{{\omega^\prime}}
\def\Om{\Omega}
\def\la{\lambda}
\def\La{\Lambda}
\def\al{\alpha}
\newcommand{\ov}{\overline}
\renewcommand{\to}{\rightarrow}
\renewcommand{\vec}[1]{\mathbf{#1}}
\newcommand{\vect}[1]{\mbox{\boldmath$#1$}}
\def\tm{{\widetilde{m}}}
\def\mcirc{{\stackrel{o}{m}}}
\newcommand{\Dm}{\Delta m}
\newcommand{\dm}{\varepsilon}
\newcommand{\tanb}{\tan\beta}
\newcommand{\nbar}{\tilde{n}}
\newcommand\PM[1]{\begin{pmatrix}#1\end{pmatrix}}
\newcommand{\up}{\uparrow}
\newcommand{\down}{\downarrow}
\def\omE{\omega_{\rm Ter}}
%

\newcommand{\Dsusy}{{susy \hspace{-9.4pt} \slash}\;}
\newcommand{\DCP}{{CP \hspace{-7.4pt} \slash}\;}
\newcommand{\mc}{\mathcal}
\newcommand{\gr}{\mathbf}
\renewcommand{\to}{\rightarrow}
\newcommand{\gtc}{\mathfrak}
\newcommand{\wh}{\widehat}
\newcommand{\br}{\langle}
\newcommand{\kt}{\rangle}

\newcommand{\Pl}{{\mbox{\tiny Pl}}}
\newcommand{\stat}{{\mbox{\tiny stat}}}
\newcommand{\tot}{{\mbox{\tiny tot}}}
\newcommand{\sys}{{\mbox{\tiny sys}}}
\newcommand{\GW}{{\mbox{\tiny GW}}}
\newcommand{\ny}[1]{\textcolor{blue}{\it{\textbf{ny: #1}}} }


\def\lsim{\mathrel{\mathop  {\hbox{\lower0.5ex\hbox{$\sim$}
\kern-0.8em\lower-0.7ex\hbox{$<$}}}}}
\def\gsim{\mathrel{\mathop  {\hbox{\lower0.5ex\hbox{$\sim$}
\kern-0.8em\lower-0.7ex\hbox{$>$}}}}}

\def\nn{\\  \nonumber}
\def\de{\partial}
\def\brf{{\mathbf f}}
\def\bbf{\bar{\bf f}}
\def\bF{{\bf F}}
\def\bbF{\bar{\bf F}}
\def\bA{{\mathbf A}}
\def\bB{{\mathbf B}}
\def\bG{{\mathbf G}}
\def\bI{{\mathbf I}}
\def\bM{{\mathbf M}}
\def\bY{{\mathbf Y}}
\def\bX{{\mathbf X}}
\def\bS{{\mathbf S}}
\def\bb{{\mathbf b}}
\def\bh{{\mathbf h}}
\def\bg{{\mathbf g}}
\def\bla{{\mathbf \la}}
\def\bmu{\mathbf m }
\def\by{{\mathbf y}}
\def\bmu{\mbox{\boldmath $\mu$} }
\def\bsig{\mbox{\boldmath $\sigma$} }
\def\bunity{{\mathbf 1}}
\def\cA{{\cal A}}
\def\cB{{\cal B}}
\def\cC{{\cal C}}
\def\cD{{\cal D}}
\def\cF{{\cal F}}
\def\cG{{\cal G}}
\def\cH{{\cal H}}
\def\cI{{\cal I}}
\def\cL{{\cal L}}
\def\cN{{\cal N}}
\def\cM{{\cal M}}
\def\cO{{\cal O}}
\def\cR{{\cal R}}
\def\cS{{\cal S}}
\def\cT{{\cal T}}
\def\eV{{\rm eV}}
%

\title{Can we probe Planckian corrections at the horizon scale with gravitational waves?}

\author{Andrea Addazi$^1$}\email{andrea.addazi@qq.com}
\author{Antonino Marcian\`o$^1$}\email{marciano@fudan.edu.cn}
\affiliation{$^1$ Center for Field Theory and Particle Physics \& Department of Physics, Fudan University, 200433 Shanghai, China}
\author{Nicol\'as Yunes$^2$}\email{nyunes@physics.montana.edu}
\affiliation{eXtreme Gravity Institute, Department of Physics, Montana State University, Bozeman, MT 59717, USA}

\begin{abstract}
\noindent
Future detectors could be used as a {\it gravitational microscope} to probe the horizon structure of merging black holes with gravitational waves. But can this microscope probe the quantum regime? We study this interesting question and find that (i) the error in the distance resolution is exponentially sensitive to errors in the Love number, and (ii) the uncertainty principle of quantum gravity forces a fundamental resolution limit. Thus, although the gravitational microscope can distinguish between black holes and other exotic objects, it is resolution limited well above the Planckian scale. 

\end{abstract}

\maketitle

\noindent 
{\emph{Introduction}}.~The recent discovery of gravitational waves (GWs)~\cite{Abbott:2016blz,Abbott:2016nmj,Abbott:2017oio,Abbott:2017vtc,Abbott:2017gyy} has raised some new and interesting ideas in fundamental black hole (BH) physics. From the possibility to observe parity violation in gravity inspired by quantum gravity~\cite{Alexander:2007kv,Yunes:2010yf,Alexander:2017jmt}, to measuring corrections to the dispersion relation~\cite{Mirshekari:2011yq}, GWs are becoming an important probe of fundamental physics~\cite{Yunes:2016jcc}. One of the ultimate fundamental questions one would like to answer relates to the full theory of quantum gravity. What is the best framework for unification? How are the Einstein equations corrected at the Planck scale? It is natural then to ask whether GWs could inform us about these questions~\cite{Jenkins:2018ysa,Calmet:2018rkj}, as future detectors become more sensitive.

In a set of pioneering studies, it was recently shown that future observations of GWs could be used to distinguish between BHs and other exotic compact objects (ECOs), i.e.~BH mimickers that do not possess a horizon~\cite{Wade:2013hoa,Cardoso:2017cfl,Sennett:2017etc,Maselli:2017cmm}. The main idea is that as compact objects coalesce, the tidal Love number imprints on the GWs emitted. Therefore, given a sufficiently sensitive observation, one could extract this tidal Love number and determine whether it is compatible with that of a BH with zero Love number or that of a horizonless object with non-zero Love number. One can thus think of GWs acting as a {\it gravitational microscope} of near horizon physics.

We here build on this work and ask whether the observation of the tidal Love number of an ECO could reveal information about the structure of spacetime with Planckian resolution. That is, we wish to determine whether the gravitational microscope can achieve Planckian resolution of near horizon physics. In particular, we explore in detail two potential limitations. First, we consider how the statistical error in the measurement of the tidal Love number propagates into error on the extraction of near horizon physics. Second, we study whether or not the uncertainty principle of quantum gravity can be evaded with the gravitational microscope.   

\vspace{0.2cm}
\noindent 
{\emph{The Love number and the gravitational microscope}}.~Consider a binary with masses $m_{1}$ and $m_{2}$ in the inspiral phase. This system can be modeled in post-Newtonian (PN) theory~\cite{Blanchet:2013haa}, a weak-field/slow-velocity expansion of the field equations, provided the two objects are sufficiently far from each other, so that non-linear relativistic corrections can be treated perturbatively. In this regime, one can safely model the response of an interferometer to an impinging GW in the frequency domain as 
 \be{aaa}
 \tilde{h}(f)=\mathcal{A}(f)e^{i\psi_{1}(f)+ i\psi_{2}(f)+i\psi_{3}(f)}\,,
 \ee
where $f$ is the GW frequency, $\mathcal{A}(f)$ is the GW Fourier amplitude, $\psi_{1}(f)$ is the contribution to the GW Fourier phase when treating the objects as spinning test particles, $\psi_{2}(f)$ is the contribution due to tidal heating, and $\psi_{3}(f)$ is the contribution due to their tidal deformability. 

Let us focus on this last contribution. To leading PN order, one can show that this contribution is
\be{psi3}
\psi_{3}(f)=-\psi_{N}\frac{\Lambda}{6m^{5}}v^{10}\frac{(1+q)^{2}}{q}\,,
\ee
where $v=(\pi m f)^{1/3}$ is the velocity, with $m=m_{1}+m_{2}$ the total mass, $\psi_{N} = (3/128) \eta^{-1} v^{-5}$ is the leading part of $\psi_{1}$, with $\eta = m_{1} m_{2}/m^{2}$ the symmetric mass ratio, and 
\be{La}
\Lambda=(1+12/q)m_{1}^{5}k_{1}+(1+12q)m_{2}^{5}k_{2}\,,
\ee
with $k_{1,2}$ the ($\ell=2$, electric-type) tidal Love numbers and $q=m_{1}/m_{2}$ the mass ratio. For two compact objects of the same type and the same mass, then $\Lambda= 26 M^{5} k$ where $k_{1} \equiv k \equiv k_{2}$ and $m_{1} \equiv M \equiv m_{2}$.

The Love number depends on the internal structure of the compact object. For the BHs of GR, the Love number vanishes~\cite{Binnington:2009bb,Landry:2014jka}, although this does not mean the horizon does not deform~\cite{Poisson:2005pi,Poisson:2009qj}. Compact objects that are not BHs, however, do have a non-zero Love number. Neutron stars, for example, have Love numbers of ${\cal{O}}(10^{2})$ depending on their equation of state~\cite{Damour:1991yw,Damour:2009vw,TheLIGOScientific:2017qsa,Abbott:2018exr}, while the boson stars so far constructed~\cite{B1,B2,B3,B4} have Love numbers of ${\cal{O}}(10)$~\cite{Cardoso:2017cfl}. ECOs, on the other hand, have Love numbers that can scale as $1/|\log(\delta)|$, where $\delta = r_{0} - r_{H}$, with $r_{0}$ the location of the ECO's surface and $r_{H}$ the location of the horizon if the ECO had been a BH of mass $M$. 

Given this, can the Love number be measured accurately enough to distinguish between a BH coalescence (which would have $\Lambda = 0$) from an ECO coalescence (which would have $\Lambda \neq 0$ but possibly small)~\cite{Wade:2013hoa,Cardoso:2017cfl,Sennett:2017etc,Maselli:2017cmm}? A Fisher analysis assuming GW detections of comparable mass binaries by LISA~\cite{LISA} suggests that this is possible. More specifically, an ECO coalescence with Love numbers of ${\cal{O}}(10^{-2})$ could be measured with a statistical accuracy of $10\%$--$50\%$~\cite{Maselli:2017cmm} using highly-spinning ``golden binaries,'' i.e.~the cleanest and loudest signals observed. This analysis employed multiple approximations, but they should be well-justified for golden binaries. Therefore, one concludes that GWs can be used as a gravitational microscope to distinguish between BHs and ECOs, provided the latter have a sufficiently large Love number.  

\vspace{0.2cm}
\noindent 
{\emph{Resolving near horizon structure}}.~Given a GW measurement of the Love number of an ECO, can one infer additional near horizon physics? Since the ECO Love number $k \propto 1/|\log(\delta)|$, can $\delta$ be inferred given a measurement of $k$? Inverting the $k$-$\delta$ relation, one finds that 
\be{delta}
\delta=r_{0}-r_{H} = r_{H} e^{-1/k}\,,
\ee
which then suggests that a measurement of $k$ and of the mass $M$ of the ECO, which determines $r_{H}$ via $r_{H} = 2 G M$ if the ECO is not spinning, yields a measurement of $\delta$. For most of the remainder of this note, we assume Eq.~\eqref{delta} is valid, but this assumption is not obvious and we will return to it in the discussion section. 
 
Let us pause for a second to scrutinize the conclusion above. The quantity $\delta$ as defined above is coordinate dependent, and thus, it is not clear whether it is observable. One possibility is to declare that $\delta$ is indeed a physical quantity related to some fundamental scale in the quantum gravity theory that leads to Eq.~\eqref{delta}. A perhaps better possibility is to think of this quantity as a proxy for an invariant measure of length, such as one constructed from a curvature invariant. For example, if one uses the Kretchmann invariant, one can construct the curvature measure ${\cal{R}} = (R_{\mu \nu \alpha \beta} R^{\mu \nu \alpha \beta})^{-1/6}$, which then yields 
\be{delta2}
\frac{{\cal{R}}_{0}}{{\cal{R}}_{BH}} 
\approx \left(\frac{M^{2}}{r_{0}^{6}}\right)^{-1/6} \left(M^{4}\right)^{-1/6} = \frac{r_{0}}{M}\,,
\ee
in a specific coordinate system where $r_{0}$ is the ECO surface and $M$ is its mass. This idea is appealing because if the quantity $\delta$ determines the quantity $k$, the latter of which is observable through its imprint on GWs, then $\delta$ ought to be describable in terms of invariant quantities. 

Given the ECO mapping between $k$ and $\delta$ in Eq.~\eqref{delta}, how small a value of $\delta$ can be inferred from a measurement of $k$? As mentioned earlier, LISA has been projected to measure $k \sim 10^{-2}$, given a supermassive BH binary signal~\cite{Maselli:2017cmm}. Equation~\eqref{delta} then implies one can infer near-horizon physics to lengths of ${\cal{O}}(M e^{-100})$, which for supermassive ECO coalescences yields lengths of ${\cal{O}}(10^{-35} \, {\rm{meters}})$ for $M = 10^{6} M_{\odot}$. Such a resolution is of ${\cal{O}}(\ell_{\Pl})$, where $\ell_{\Pl}$ is the Planck length, near the ECO surface. Pushing this further, similar observations with lower mass binaries or higher signal-to-noise ratio, using e.g.~U-DECIGO \cite{UDECIGO}, BBO \cite{BBO}, Tian-Qing \cite{TQ} or TAIJI \cite{TAIJI}, would probe sub-Planckian distances. 

\vspace{0.2cm}
\noindent 
{\emph{Statistical uncertainty in measurements of near horizon structure}}. We now consider the accuracy to which the length difference $\delta$ can be measured, given a measurement of the Love number $k$. The best-fit value of $k$ and of the ECO masses possess statistical uncertainty. The latter can be estimated from the diagonal components of the variance-covariance matrix $\Sigma^{ab}$, where the superscripts run over the model parameters $\lambda^{a} = (M,k)$. The $\Sigma^{ab}$ matrix can be estimated through a Fisher analysis, as we explain in the appendix, or alternatively obtained from a Markov-Chain Monte-Carlo exploration of the likelihood probability distribution in a Bayesian analysis. 

Given the variance-covariance matrix, simple error propagation can be used to find the statistical uncertainty in the inferred parameter $\delta$, namely  
\begin{align}
\label{eq:Delta-error-pre}
\sigma^{\stat}_{\delta} = \sqrt{\left(\frac{d \delta}{d\lambda^{a}}\right) \left(\frac{d \delta}{d\lambda^{b}}\right) \Sigma^{ab}}\,.
\end{align}
Using the $k$-$\delta$ mapping for ECOs in Eq.~\eqref{delta}, this yields
\begin{align}
\label{eq:Delta-error-pre-2}
\frac{\sigma^{\stat}_{\delta}}{\hat{\delta}} &= \sqrt{\left(\frac{\sigma^{\stat}_{M}}{\hat{M}}\right)^{2} + \frac{1}{\hat{k}^{2}} \left(\frac{\sigma^{\stat}_{k}}{\hat{k}}\right)^{2}}\,,
\end{align}
where $\hat{\lambda}^{a} = (\hat{M},\hat{k})$ are the best-fit values of the parameters, and we have neglected covariances between the mass and the Love number. We then see clearly that for small measurements of $k$, the statistical uncertainty in $\delta$ scales as $\sigma^{\stat}_{\delta} \propto 1/\hat{k}$. The value of the measured Love number at which the statistical uncertainty equals the inferred values of $\hat{\delta}$ is $\hat{k} \approx 0.2$ for the expected statistical accuracy in the estimation of $M$ and $k$. Thus, any inferred value of $\hat{\delta}$ derived from $\hat{k} < 0.2$ will be dominated by statistical uncertainty because the uncertainty in $\hat\delta$ is exponentially affected by the uncertainty in $\hat k$. 

\vspace{0.2cm}
\noindent 
{\emph{Systematic uncertainty in measurements of near horizon structure}}. Even if one could distinguish the ECO radius from the corresponding Schwarzschild radius with Planckian precision, one would be in the resolution limit $\delta \rightarrow \ell_{\Pl}$. As a consequence of the quantum uncertainty principle, this corresponds to a momentum resolution 
\begin{align}
\Delta p>\hbar \; \ell_{\Pl}^{-1}= \hbar \sqrt{c^{3}/G\hbar}\,,
\end{align}
which implies a huge uncertainty in the four-momenta of the inspiraling bodies, including both their rest mass and their inspiral velocities. This, in turn, implies a fundamental quantum uncertainty in the interaction or binding energy of the two objects, and thus, in the acceleration of the bodies and the GWs they emit.

Let us pause again to scrutinize the above argument. Lacking a complete quantum gravity theory, one may argue that perhaps the uncertainty principle should not apply here. Spacetime, however, is defined by a manifold, which by definition reduces to flat spacetime in a small neighborhood about any point. The 2-sphere that defines the location of the ECO surface is not special, and curvature effects are relatively weak on it for supermassive objects, as the curvature scales inversely with the mass. Therefore, one can choose any point on or near this 2-sphere and consider a small neighborhood about it larger than the Planck length, in which spacetime will look flat. In this neighborhood, quantum principles, like the uncertainty principle, should continue to hold.

The percolation of quantum uncertainty into GWs implies a fundamental limitation in the accuracy to which any GW model parameter can be extracted because the signal becomes quantum fuzzy. For example, in the $\delta \rightarrow \ell_{\Pl}$ limit, the uncertainty in the mass $\Delta M\rightarrow \sqrt{\hbar c/G}$, which corresponds to the Planck scale. This, in turn, percolates into the gravitational interaction, since the uncertainty in the gravitational binding energy $\Delta E_{b}\rightarrow -G \eta m \Delta M/r$ at leading PN order. But the binding energy affects the rate at which the orbital and the GW frequency changes via the balance law
\begin{align}
\frac{dF}{dt} = \left(\frac{dE_{b}}{dt}\right) \left(\frac{dE_{b}}{dF}\right)^{-1} = -\left(\frac{dE_{\GW}}{dt}\right) \left(\frac{dE_{b}}{dF}\right)^{-1}\,,
\end{align}
where $F$ is the orbital frequency and $dE_{\GW}/dt$ is the rate at which energy is removed from the system by GW emission. Therefore, quantum uncertainty in the binding energy translates into a quantum uncertainty in the orbital frequency, which then percolates into an uncertainty in the GW frequency and its phase of the signal itself, preventing measurements beyond the Planck scale. 

What is the fundamental limitation that quantum uncertainty in the signal places on the accuracy to which model parameters can be fitted? Quantum fluctuations in the signal of ${\cal{O}}(\ell_{\Pl})$ blur or fuzz out its amplitude and phase, and so when one fits this quantum fuzzy signal to waveform templates, the accuracy to which model parameter can be estimated will be limited by a systematic uncertainty of the same size as the quantum jitter itself. The total uncertainty in the extraction of any parameter in a waveform model is then the sum of the statistical error $\sigma_{\stat}$ (described in Eq.~\eqref{eq:Delta-error-pre-2}) and a systematic error $\sigma_{\sys} = {\cal{O}}(\ell_{\Pl})$ in quadrature, leading to 
\begin{align}
\label{eq:Delta-error-pre-3}
\frac{\sigma_{\delta}^{\tot}}{\hat{\delta}} &= \sqrt{\left(\frac{\sigma^{\stat}_{M}}{\hat{M}}\right)^{2} + \frac{1}{\hat{k}^{2}} \left(\frac{\sigma^{\stat}_{k}}{\hat{k}}\right)^{2} + \frac{a^{2} \ell_{\Pl}^{2}}{\hat{\delta}^{2}}}\,,
\end{align}
where we have set $\sigma_{\sys} = a \; \ell_{\Pl}$ for $a \in \Re$ and of ${\cal{O}}(1)$. Clearly then, quantum uncertainty provides a floor for the uncertainty in the measurement of $\delta$, as we can see in Fig.~\ref{fig:lpl}. Observe that the total $1\sigma$ error is ${\cal{O}}(10^{2})$ times larger than the inferred value, saturating at the quantum uncertainty floor at high spins. This saturation would occur at lower spin values if we had chosen a smaller variance for the $\hat{k}$ measurement. 
\begin{figure}
        \includegraphics[width=0.54\textwidth,clip=true]{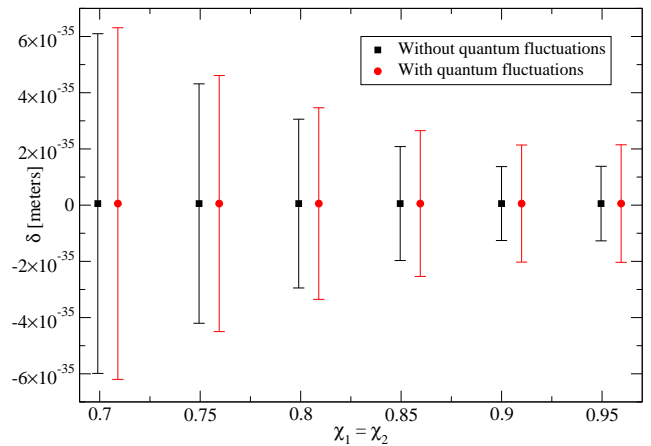}
        \caption{(Color Online). Inferred value of $\delta$ and $1\sigma$ errors with (black, Eq.~\eqref{eq:Delta-error-pre-2}) and without (red, Eq.~\eqref{eq:Delta-error-pre-3}) quantum fluctuations. We have here assumed a GW observation of a compact ECO inspiral with $m_{1} = 1.1\times 10^{6} M_{\odot}$ and $m_{2} = 10^{6} M_{\odot}$, dimensionless spin $\chi_{1,2}$, Love numbers $k_{1} = k_{2} = 0.02$ at a distance of 2Gpc. For the uncertainties in $M$ and $k$, we used $\sigma_{M}^{\stat}/\hat{M} = 10^{-5}$ and $\sigma_{k}^{\stat}/\hat{k} \in (0.2,1)$ (see the appendix), while for the quantum fluctuations we set $a = 1$. Observe that the error bars on the inferred value of $\delta$ are much larger than the measurement itself.}
        \label{fig:lpl}
\end{figure}

\vspace{0.2cm}
\noindent 
{{\emph{Discussion.}}~We have investigated the resolving power of the gravitational microscope to use a GW measurement of the Love number $k$ to infer near horizon physics a distance $\delta$ from an ECO surface. For future observations with LISA, we have found that the resolution in $\delta$ is limited by statistical error when $\hat{k} < 0.2$. In particular, the statistical error is only controlled if the statistical uncertainty in the Love number is less than the squared of the inferred Love number, as shown in Eq.~\eqref{eq:Delta-error-pre-2}. For a measurement of the Love number of $\hat{k} = 10^{-2}$, this implies a fractional accuracy better than $\sigma_{k}^{\stat}/\hat{k} = 1\%$, which would require signal-to-noise ratios above $10^{4}$.   

The above statistical considerations neglect the effect of systematics in the modeling of the waveform itself. All models used to date assume compact binaries in isolation (i.e.~in vacuum), but supermassive compact object binaries may have a circumbinary accretion disk, or they may be perturbed by a third body. The presence of such effects could impact the GW signal~\cite{Yunes:2011ws,Kocsis:2011dr,Barausse:2014tra,Yunes:2010sm} and, the use of vacuum waveforms to fit this signal could incorrectly lead to non-zero measurements of Love numbers, which could be in turn incorrectly associated with ECOs. Mismodeling error~\cite{Cutler:2007mi} is a form of systematic uncertainty that becomes more severe for high signal-to-noise ratio events and must thus be included in the total error budget, which could further limit inferences made about ECOs from Love number measurements. 

Putting mismodeling systematics aside, we have also found that the resolving power of the gravitational microscope will also be limited by systematics in the signal due to quantum fluctuations. If the uncertainty principle of quantum mechanics is valid near the horizon of ECOs, then quantum fluctuations in the four-momenta of the objects will percolate into a systematic uncertainty in the amplitude and phase of the GW signal. We have estimated this uncertainty to be of order the Planck scale, but in principle it could be larger, for example of order the string length, since typically the hierarchy between these scales is governed by the compactification volume and the string coupling. A better understanding of quantum gravity, for example through the completion of quantum gravity theories and the numerical study of the coalescence of quantum gravity compact objects, could aid in quantifying more precisely the impact of quantum fluctuations in GW observables.  

But if quantum fluctuations are truly present in the gravitational measurement of distances at the Planck scale, then sub-Planckian measurements ought to be impossible. From the effective quantum gravity framework, at such scales quantum fluctuations become uncontrollable and one losses the very concept of spacetime continuity with the emergence of spacetime foaminess. From the quantum field theory perspective, this is related to the non-renormalizability of the theory, and at sub-Planckian scales one expects the emergence of different virtual spacetime topologies --- for example virtual BH pairs that create and annihilate. Because of this, the very notion of a classical BH horizon as a Cauchy surface loses meaning at the Planck scale. 

Unfortunately, the current status of quantum gravity models prevents us from going any further in this line of questioning. Without a full model, even the construction of an isolated compact object with quantum gravity corrections is missing. In this paper, we have studied the possibility of using the ECO relation between $k$ and $\delta$ in Eq.~\eqref{delta} to see if a measurement of the former allows for microscopic measurements of the latter, but it is unclear whether this relation persists in quantum gravitational compact objects. The relation has only been shown to hold for wormholes~\cite{W} and gravastars~\cite{G1,G2}, which as~\cite{Yunes:2016jcc} put it are both examples of cut-and-paste metrics: wormholes are Schwarzschild metrics glued together at a finite radius, while gravastars are an exterior Schwarzschild metric glued to an interior de Sitter metric. To our knowledge, neither of these classical metrics arises as a solution to a quantum gravity model, they have not been shown to arise generically from gravitational collapse, and even if they did, they would be at least unstable when spin is included, unless the ECO's surface is somehow sufficiently absorbing~\cite{Hod:2017cga,Maggio:2018ivz}. 

These observations then suggest the question of whether the exponential relation between $k$ and $\delta$ is also realized in other spacetimes for compact objects with quantum-gravity inspired modifications. Several insightful attempts have been made to construct such objects, e.g.~fuzzball string condensates \cite{Mathur:2005zp,Guo:2017jmi,Bianchi:2017bxl}, graviton condensates \cite{Dvali:2012en,Addazi:2016ksu,Ciafaloni:2017ort}, or string holes \cite{Veneziano:2012yj}. Alternative formulations of non-perturbative quantum gravity BHs also exist~\cite{Bojowald:2012ux,Amelino-Camelia:2016gfx,BenAchour:2016brs}, as well as BH solutions in effective field theory expansions of heterotic string theory, as in Einstein-dilaton-Gauss-Bonnet gravity~\cite{Kleihaus:2003sh,Yunes:2011we,Kleihaus:2011tg} and dynamical Chern-Simons gravity~\cite{Alexander:2009tp,Yunes:2009hc} (see appendix).  None of these is perfect, and in fact, they all have model-specific problems. But what they do have in common is that  the $k$-$\delta$ mapping in Eq.~\eqref{delta} either does not hold (because the Love number vanishes~\cite{Kleihaus:2003sh,Yunes:2011we,Kleihaus:2011tg,Alexander:2009tp,Yunes:2009hc}), or is not expected to hold at Planck scales. It is thus unclear how the subset of ECOs for which Eq.~\eqref{delta} holds is connected to compact objects with quantum gravity modifications. 

This discussion then brings us back to the \emph{generality} of the $k$-$\delta$ mapping in Eq.~\eqref{delta}. Even if the quantum uncertainty and the statistical issues were not present, the ability to agnostically probe Planckian distances with a measurement of the Love number depends strongly on the validity of this mapping. As explained above, the mapping is not general since there exist counter-examples: some BHs in modified gravity have zero Love number (and thus no $k$-$\delta$ mapping)~\cite{Kleihaus:2003sh,Yunes:2011we,Kleihaus:2011tg,Alexander:2009tp,Yunes:2009hc}}, and some ECOs, like the boson stars so-far constructed, have a different $k$-$\delta$ mapping~\cite{Cardoso:2017cfl}. The $k$-$\delta$ mapping in Eq.~\eqref{delta} then seems to hold for a very special subset of compact objects, which introduce deviations from BH spacetimes that are not necessarily mediated by curvature corrections, but are rather generated through cut-and-paste procedures. Since the $k$-$\delta$ mapping in Eq.~\eqref{delta} is only valid for such specific models, the assumption that it is also valid for generic compact objects that arise in quantum gravity is very strong and not necessarily well-justified.   


\emph{Acknowledgments} -- We would like to thank Stephon Alexander, Emanuele Berti, Neil Cornish, Luis Lehner, Tyson Littenberg, Frans Pretorius, Carlo Rovelli, and Thomas Sotiriou for their encouragement and useful comments when writing this note. AA and AM are also grateful to Abhay Ashtekar, Massimo Bianchi, and Thibault Damour for useful discussions on these subjects. We also would like to thank Vitor Cardoso, Andrea Maselli and Paolo Pani for other discussions. AA and AM wish to acknowledge support by the NSFC, through the grant No. 11875113, the Shanghai Municipality, through the grant No. KBH1512299, and by Fudan University, through the grant No. JJH1512105. NY acknowledges support from NSF grant PHY-1759615, NSF  and NASA grants NNX16AB98G and 80NSSC17M0041.


\section{Appendix}
\label{sec:appendix}

\subsection{Fisher analysis}
\label{FA}
\noindent 
The variance-covariance matrix can be estimated through a Fisher analysis by computing the inverse of the Fisher matrix, which in turn is given by the noise-weighted inner product of the $\lambda^{a}$-directional derivative of the Fourier transform of the waveform. The off-diagonal elements of this matrix represent the correlation between the parameters, while the diagonal elements give the variance squared, namely $\Sigma^{MM} = \sigma_{M}^{2}$,  $\Sigma^{kk} = \sigma_{k}^{2}$. The variance for the mass parameter can be obtained from the variance of the chirp mass and the symmetric mass ratio, computed for example in~\cite{Berti:2004bd}, yielding $\sigma_{M}/M \in (0.005,0.015)$ for supermassive BH inspirals at a $3$Gpc distance and detectable with LISA at SNRs of roughly $2000$. The variance of the Love number can be read directly from~\cite{Maselli:2017cmm}, yielding $\sigma_{k}/k = \sigma_{\Lambda}/\Lambda \in (0.2,1)$.

In the analysis presented in the note that led to Eq.~\eqref{eq:Delta-error-pre-2}, we neglected any correlations between $M$ and $k$. That is, we used the fact that generally $\Sigma^{kk} \gg \Sigma^{Mk} \ll \Sigma^{MM}$ to ignore $\Sigma^{Mk}$ in the error propagation calculation. This is a justified approximation because $M$ is encoded in the early inspiral part of the waveform, while $k$ is encoded in the very late inspiral. Moreover, including this correlation would only make the statistical error in Eq.~\eqref{eq:Delta-error-pre-2} larger, and thus, it would not affect the arguments presented in the note.  

With that in mind, we can now compute the variance of the inferred $\delta$ measurement, given measurements of the Love number and the mass to a certain statistical accuracy. Using Eq.~\eqref{delta} in Eq.~\eqref{eq:Delta-error-pre}, we find
\begin{align}
\label{eq:Delta-error-2}
\sigma_{\delta}^{2} &= 
4 e^{-2/k} \sigma_{M}^{2}
+ \frac{4 M^{2}}{k^{2}} e^{-2/k} \left(\frac{\sigma_{k}}{k}\right)^{2}\,,
\end{align}
which can then be rewritten as in Eq.~\eqref{eq:Delta-error-pre-2}. Both terms in this expression are similarly large, but clearly the second term dominates in the $k\ll1$ limit. 

Two main features clearly arise from the above expression. The first is that for sufficiently small $k$, such that $\hat\delta$ approaches the Planck regime, the variance in $\hat\delta$ grows as $k^{-2}$. This is a result of the non-linear exponential mapping between $\delta$ and $k$ in Eq.~\eqref{delta}: even a tiny error in $\hat{k}$ can lead to a huge error in $\hat\delta$. 

What is not included in the analysis above is the systematic error in the measurement of $\delta$ induced by quantum fluctuations, which we presented in Eq.~\eqref{eq:Delta-error-pre-3} of the note. From this equation, it is clear that even if one ignores the variance in $\hat{k}$, the variance in $\hat\delta$ eventually saturates at $\ell_{\Pl}$, preventing any measurements of $\delta$ below the Planck scale. This must be the case, as otherwise sufficiently precise measurements of $k$ would enable probes above Planckian scales.

\subsection{Euclidean Quantum Gravity}
\label{EQG}
\noindent 
In quantum gravity, the metric tensor $g_{\mu\nu}$ is promoted to a quantum operator\footnote{In this appendix, the overhead hat notation stands for an operator quantity and not for the best fit value in parameter estimation, as it does in the rest of the note.} $\hat{g}_{\mu\nu}$ that depends functionally on the space-time field. In an effective quantum field theory framework, this operator can be separated into a background metric and a graviton quantum field operator
$$\hat{g}_{\mu\nu}=\bar{g}_{\mu\nu}+\hat{h}_{\mu\nu}\,.$$ 
The quantum fluctuations of the metric enter into the definition of quantum distance via $ds=\sqrt{g_{\mu\nu}dx^{\mu}dx^{\mu}}$, and as a consequence, the $\delta$ parameter must be promoted to a quantum operator at Planckian separations, i.e. 
\be{delta3}
\langle \hat{\delta} \rangle=\langle \hat{r}_{0}-2\hat{M} \rangle\,.
\ee
In the path integral formulation of quantum gravity, the expectation value of $\hat{\delta}$ is given by
\be{delta2}
\langle \hat{\delta}\rangle=\langle vac|\, \hat{\delta}\, |vac\rangle=\int \mathcal{D}g_{\mu\nu}\, \mathcal{D}\Psi \, e^{iS[g,\Psi]}\,\hat{\delta}(g,\Psi)\, , 
\ee
where $\langle vac|\, \hat{\delta}\, |vac\rangle$ is the quantum gravity path integral of the $\hat{\delta}$ operator, defined as an expectation value on the vacuum state of the theory, $\mathcal{D}(...)$ are the usual Feynman path integral measures on the metric and the matter fields $\Psi=\{\psi_{i},\,\phi_{a},\,A_{i'}\}$ (fermion, scalar and boson species respectively), and $S[g,\Psi]$ is the effective quantum gravity action coupled to quantum matter fields. 

In the classical limit, the correspondence principle of quantum mechanics guarantees the good convergence of the $\hat{\delta}$ operator to the classical notion of distance. That is, in the large wavelength $\lambda$ limit ($\lambda \gg \ell_{\Pl}$),
\be{llll}
\langle \hat{\delta}\rangle \rightarrow \delta +O(\ell_{\Pl})\,,
\ee
when $\delta \gg \ell_{\Pl}$. But if $\delta \sim \ell_{\Pl}$, the quantum gravity uncertainty principle prevents us from distinguishing $r$ from $2M$ with Planckian precision because of the $O(\ell_{\Pl})$ fluctuations of the expectation value of the $\hat{\delta}$ operator. 

In the $\delta \gg \ell_{\Pl}$ limit, a semiclassical quantum gravity approach can be applied to find the corrections to Eq.~\eqref{llll} by integrating on a background metric -- for example, the Schwarzschild metric -- and considering the graviton field fluctuations around it. The quantum gravity action is then just the Einstein-Hilbert action in the presence of boundary terms, coupled to matter and linearized. In the Euclidean partition function, this corresponds to the linearized version of 
\begin{align}
\label{EinsteinHilbert}
S_{EH}&=-\int_{\Sigma}\!\!\sqrt{g}\, d^{4}x\, \Big( \mathcal{L}_{m}+\frac{1}{16\pi G}R \Big)
\nonumber \\
&+\frac{1}{8\pi G}\int_{\partial \Sigma}\!\!\sqrt{g'}\,d^{3}x\, (K-K^{0})\, ,
\end{align}
where $\mathcal{L}_{m}$ is the matter Lagrangian, $K$ is the trace of the extrinsic curvature induced on the boundary of the integration region $\Sigma$, $g'$ is the induced metric on the boundary $\partial \Sigma$, and $K^{0}$ is the trace of the extrinsic curvature embedded in  flat spacetime. Linearizing this action via $\Psi=\Psi_{0}+\delta \Psi$ and $g=\bar{g}+h$,
\be{Li}
S[\Psi,g]=S_{0}+S_{2}[\delta \Psi]+S_{2}[h]+O\{(\delta \Psi)^{2},(h)^{2}\}\,,
\ee
where $S_{0}\equiv S[\Psi_{0},\bar{g}]$.  The semiclassical-limit of the $\langle \hat{\delta}\rangle$ can be then split in a classically defined quantity and a quantum fluctuating part, i.e. 
\be{delta2}
\langle \hat{\delta} \rangle\rightarrow \delta + d_{2}\delta \,,
\ee
where $\delta$ corresponds to the $S_{0}$ classical action, while $d_{2}\delta$ corresponds to the $S_{2}$-part, containing the linearized fluctuation. 

Quantum gravity loop-corrections are $O(\alpha_{G}^{n}\hbar^{n})$, where $\alpha_{G} \propto G_N E^{2}$ is the gravitational coupling, $G_N$ is Newton's constant, $E$ is the center of mass energy, and $n$ is the number of vertices in loops. These loop-corrections generate new effective higher-order derivative correlators of gravitons with respect to the Einstein-Hilbert tree-level action. Thus, the action is deformed, by radiative corrections, to an effective quantum gravity action that has the quadratic gravity structure~\cite{Yunes:2011we} 
\be{EFTQG}
\int \!\!d^{4}x \sqrt{-g}(R\!+\!c_{1}R^{2}\!+\!c_{2}R_{\mu\nu}R^{\mu\nu}\!+\!c_{3}R_{\mu\nu\rho\sigma}R^{\mu\nu\rho\sigma}+...)\,.
\ee
The $c_{i}$ coefficients cannot be controlled at the Planck scale, and thus, the effective action may turn into a non-local one when an infinite number of loops is taken into account. The uncertainty on the coefficient $c_{i}$ at the Planck scale then percolates into an unpredictable expectation value $\langle \hat{\delta}\rangle$.

\end{document}